\documentclass[10pt]{revtex4}
\usepackage{amssymb}
\usepackage{latexsym}
\usepackage{epsfig}
\usepackage{amsmath}

\begin{document}
\title{Holographic dark energy in the DGP braneworld with GO cutoff}
\author{S. Ghaffari$^{1}$, M. H. Dehghani$^{1,2}$\footnote{mhd@shirazu.ac.ir} and A. Sheykhi$^{1,2}$
\footnote{asheykhi@shirazu.ac.ir}}
\affiliation{$^1$ Physics Department and Biruni Observatory, College of
Sciences, Shiraz University, Shiraz 71454, Iran\\
         $^2$  Research Institute for Astronomy and Astrophysics of Maragha
         (RIAAM), P.O. Box 55134-441, Maragha, Iran}

\begin{abstract}
We consider the holographic dark energy (HDE) model in the
framework of DGP braneworld with Granda-Oliveros infrared (IR)
cutoff, $L=(\alpha \dot{H}+\beta H^2)^{-1/2}$. With this choice
for IR cutoff, we are able to derive evolution of the cosmological
parameters such as the equation of state and the deceleration
parameters, $w$ and $q$, as the functions of the redshift parameter $z$. As far
as we know, most previous models of HDE presented in the
literatures, do not gives analytically $\omega=\omega(z)$ and $q=q(z)$. We
plot the evolution of these parameters versus $z$ and discuss
that the results are compatible with the recent observations. With
suitably choosing the parameters, this model can exhibit a
transition from deceleration to the acceleration around $z\approx
0.6$. Then, we suggest a correspondence between the quintessence
and tachyon scalar fields and HDE in the framework of DGP
braneworld. This correspondence allows us to reconstruct the
evolution of the scalar fields and the scalar potentials. We also
investigate stability of the presented model by calculating the
squared sound speed, $v^2_s$, whose sign determines the stability
of the model. Our study shows that $v^2_s$ could be positive
provided the parameters of the model are chosen suitably. In
particular, for $\alpha>1$, $\beta>0$, and $\alpha<1$, $\beta<0$,
we have $v^2_s>0$ during the history of the universe, and so the
stable dark energy dominated universe can be achieved. This is in
contrast to the HDE in standard cosmology, which is unstable against
background perturbations and so cannot lead to a stable dark
energy dominated universe.

\end{abstract}

\maketitle

\section{Introduction}
A complementary astrophysical data from type Ia Supernova, Large
Scale Structure (LSS) and Cosmic Microwave Background (CMB)
indicate that our Universe is currently undergoing a phase of
accelerating \cite{Riess}. A component which is responsible for
this accelerated expansion is usually dubbed "dark energy" (DE).
The simplest candidate for DE is the cosmological constant
\cite{Weinberg} which is located in the center both from
theoretical and observational evidences. However, there are
different alternative theories for the dynamical DE scenarios
which have been proposed to interpret the accelerating universe.
One of these models, which has arisen a lot of enthusiasm recently,
is HDE. It was shown by Cohen et al. \cite{Cohen} that in quantum
field theory a short distance cutoff could be related to a long
distance cutoff (IR cutoff L) due to the limit sets by black hole
formation. If the quantum zero-point energy density is due to a
short distance cutoff, then the total energy in a region of size
$L$ should not exceed the mass of a black hole of the same size,
namely $L^3\rho_D\leq L M_{pl}^2$. The largest $L$ is the one
saturating this inequality, so that we obtain the HDE density as
\cite{Cohen}
\begin{equation}\label{HDE}
\rho_D=3c^2M_{\rm pl}^2L^{-2},
\end{equation}
where $M^2_{\rm pl} = 8\pi G$ is
the reduced Planck mass. In general the factor $c^2$ in holographic
energy density can vary with time very slowly \cite{pav4}.
By slowly varying we mean that $\dot{(c^2)}/c^2$ is upper bounded
by the Hubble expansion rate, $H$, i.e., \cite{pav4}
\begin{equation}\label{cdot}
 \frac{\dot{(c^2)}}{c^2}\leq H.
\end{equation}
It was also argued that $c^2$ depends on the IR length, $L$
\cite{pav4}. For the case of $L=H^{-1}$, one can take $c^2$
approximately constant in the late time where DE dominates
($\Omega_m<1/3$) \cite{pav4}. Since in the present work we
consider the late time cosmology where the DE dominates, we shall
assume the factor $c^2$ to be constant.

Depending on the IR cutoff, $L$, many HDE models have been
proposed in the literatures. A comprehensive, but not complete,
list of IR cutoffs which have been used includes the particle
horizon radius $R_H\equiv a\int^t_0dt/a=a\int^a_0 da/(Ha^2)$
\cite{Fischler,Li}, the Hubble horizon $L=H^{-1}$
\cite{Horava,Hsu,pav1}, the future event horizon radius $R_h\equiv
a\int_t^\infty dt/a=a\int_a^\infty da/(Ha^2)$ \cite{Li}, the
apparent horizon radius $L=(H^2+k/a^2)^{-1/2}$ \cite{Shey1},   the
Ricci scalar curvature radius $R_{CC}= (\dot{H}+H^2)^{-1/2}$
\cite{Gao}, the so-called Granda-Oliveros (GO) cutoff, which is
the formal generalization of $R_{CC}$, namely $L = (\alpha
\dot{H}+\beta H^2)^{-1/2}$ \cite{Granda1,Granda2,Jamil1}, the age
of our Universe $T=\int_0^ada/(Ha)$ \cite{Cai}, the conformal age
of our Universe $\eta\equiv\int_0^t dt/a=\int^a_0da/(a^2H)$
\cite{Wei, Shey2}, the radius of the cosmic null hypersurface
\cite{Gao2}, etc.

On the other side, in recent years, the theories of large extra
dimensions in which the observed universe is realized as a brane
embedded in a higher dimensional spacetime, have received a lot of
interest.  In these theories the cosmological evolution on the
brane is described by an effective Friedmann equation that
incorporates non-trivially with the effects of the bulk onto the
brane. One of the well-know picture in the braneworld scenarios
was proposed by Dvali-Gabadadze-Porrati (DGP) \cite{DGP}. In
this model our four-dimensional Universe is a
Friedmann-Robertson-Walker (FRW) brane embedded in a
five-dimensional Minkowskian bulk with infinite size. In this
model the recovery of the usual gravitational laws on the brane is
obtained by adding an Einstein-Hilbert term to the action of the
brane computed with the brane intrinsic curvature. The
self-accelerating branch of DGP model can explain the late time cosmic
speed-up without recourse to DE or other components of energy
\cite{Def, Def1}. However, the self-accelerating DGP branch has
ghost instabilities and it cannot realize phantom divide crossing
by itself. To realize phantom divide crossing it is necessary to
add at least a component of energy on the brane. On the other
hand, the normal DGP branch cannot explain acceleration but it has
the potential to realize a phantom-like phase by dynamical
screening on the brane.

In the present work we consider the HDE model in the framework of
DGP braneword with GO cutoff, $L=(\alpha \dot{H}+\beta
H^2)^{-1/2}$, proposed in \cite{Granda2}. Our work differs from
\cite{Granda2} in that, they studied HDE model with GO cutoff in
standard cosmology, while we investigate this model in the
framework of DGP braneword and incorporate the effect of the extra
dimension on the evolution of the cosmological parameters on the
brane. {The main difference between the HDE with GO cutoff in the
framework of DGP braneword, with the one considered in standard
cosmology \cite{Granda2}, is that the equation of state parameter
of the HDE with GO cutoff in standard cosmology is a constant
\cite{Granda2}, namely
\begin{equation}
w_D=-1+\frac{2(\alpha-1)}{3\beta},
\end{equation}
however, as we shall see in DGP braneworld, due to the bulk
effects, $w_D$ becomes a time variable parameter. Clearly, a time
variable DE is more compatible with observations. In particular,
the analysis of data from WMAP9 or Planck-$2013$ results on CMB
anisotropy, BAO distance ratios from recent galaxy surveys,
magnitude-redshift relations for distant SNe Ia from SNLS3 and
Union2.1 samples indicate that the time varying DE gives a better
fit than a cosmological constant \cite{Planck1,Planck2}. In
addition, current data still slightly favor the quintom DE
scenario with EoS across the cosmological constant boundary
$w_D=-1$ \cite{Planck1}. Furthermore, we are able to derive
explicitly, the cosmological parameters on the brane as functions
of redshift parameter and provide a profile of the cosmic
evolution on the brane. Indeed, this is the advantages of the
present model in compared to all other HDE models. As far as we
know, this is the first model of HDE which leads to analytical
solution, $w=w(z)$ and $q=q(z)$.} Then, we establish the
correspondence between our model and scalar field models of DE and
reconstruct the evolution of scalar fields and potentials.
Finally, we investigate the stability of this model against
perturbation in different cases. We find that for some range of
the parameter spaces our model is stable which indicates the
viability of this model for explanation of the late time
acceleration.
\section{HDE in DGP braneworld}
We consider a homogeneous and isotropic FRW universe on the brane
which is described by the line element
\begin{equation}
{\rm d}s^2=-{\rm d}t^2+a^2(t)\left(\frac{{\rm d}r^2}{1-kr^2}+r^2{\rm
d}\Omega^2\right),\label{metric}
\end{equation}
where $k=0,1,-1$ represent a flat, closed and open maximally
symmetric space on the brane, respectively.
 The modified Friedmann equation in  DGP braneworld model is given by \cite{Def}
\begin{equation}\label{Friedeq01}
H^2+\frac{k}{a^2}=\Big(\sqrt{\frac{\rho}{3M_{\rm
pl}^2}+\frac{1}{4r^2_c}}+\frac{\epsilon}{2r_c}\Big)^2,
\end{equation}
where $\epsilon=\pm1$ corresponds to the two branches of solutions
\cite{Def}, and $r_{\rm c}$ stands for crossover length scale
between the small and large distances in DGP braneworld defined as
\cite{Def}
\begin{equation}
r_c=\frac{M_{\rm pl}^2}{2M_{5}^3}=\frac{G_5}{2G_4}.
\end{equation}
We also assume that there is no energy exchange between the brane and
the bulk and so the energy conservation equation holds on the
brane,
\begin{equation}
\dot{\rho}+3H(1+w)\rho=0.\label{Conserveq}
\end{equation}
For $r_c\gg1$, the Friedmann equation in standard cosmology is
recovered
\begin{equation}
H^2+\frac{k}{a^2}=\frac{\rho}{3M_{\rm pl}^2}.
\end{equation}
Recent observations indicate that our universe is spatially flat.
For a flat FRW universe on the brane, Eq. (\ref{Friedeq01})
reduces to
\begin{equation}\label{Friedeq02}
H^2-\frac{\epsilon}{r_c}H=\frac{\rho}{3M_{\rm pl}^2}.
\end{equation}
Depending on the sign of $\epsilon$, there are two different
branches for the DGP model. For $\epsilon=+1$ and in the absence
of any kind of energy or matter field on the brane ($\rho=0$),
there is a de-Sitter solution for Eq. (\ref{Friedeq02}) with
constant Hubble parameter
\begin{equation}\label{de sitter}
H=\frac{1}{r_c}\Rightarrow a(t)= a_{0} e^{\frac{t}{r_c}}.
\end{equation}
Clearly, Eq. (\ref{de sitter}) leads to an accelerating universe
with constant equation of state parameter $\omega_D=-1$, exactly
like the cosmological constant. However, there are some
unsatisfactory problems with this solution. First of all, it
suffers the well-known cosmological constant problems namely, the
fine-tuning and the coincidence problems. Besides, it leads to a
constant $\omega_D$, while many cosmological evidences, especially
the analysis of the type Ia supernova data indicates that the time
varying DE gives a better fit than a cosmological constant
\cite{Planck1,Planck2,Alam}. Most of these data favor the
evolution of the equation of state parameter and in particular it
can have a transition from $\omega_D>-1$ to $\omega_D < -1$ at
recent stage. Although some evidence such as the galaxy cluster
gas mass fraction data do not support the time-varying $\omega_D$
\cite{Chen}, an overwhelming flood of papers has appeared to
understand the $\omega_D=-1$ crossing in the past decade
\cite{Wang}. In addition, to arrive at Eq. (\ref{de sitter}) one
ignores all parts of energy on the brane including DE, dark matter
and byronic matter, which is not a reasonable assumption.

On the other hand, for $\epsilon=-1$ and $r_c \ll H^{-1}$, one can
neglect the term $H^2$ in Eq. (\ref{Friedeq02}) to arrive at
\begin{equation}\label{RS}
H^2=\frac{\rho^2}{36M_5^6}.
\end{equation}
This is the Friedmann equation in spatially flat RS II braneworld
\cite{RS}. Clearly, Eq. (\ref{RS})  does not have a self
accelerating solution, so it implies the requirement of some kind
of DE on the brane.

In the present paper, we consider the HDE model in flat DGP
braneworld with GO cutoff, which is defined as \cite{Granda2}
\begin{equation}\label{GO cutoff}
L=(\alpha H^2+\beta\dot{H})^{-1/2}.
\end{equation}
With this IR cutoff, the energy density (\ref{HDE}) can be written
\begin{equation}\label{rho1}
\rho_D=3M_{\rm pl}^2(\alpha H^2+\beta\dot{H}),
\end{equation}
where $\alpha$ and $\beta$ are constants which should be
constrained by the recent observational data and we have absorbed
the constant $c^2$ in $\alpha$ and $\beta$. Hereafter, we work in
a unit in which $M_{\rm pl}^2=1$. {We also restrict our study to
the current cosmological epoch, and hence we are not considering
the contributions from matter and radiation by assuming that the
dark energy $\rho_D$ dominates, thus the Friedman equation becomes
simpler.} Substituting Eq. (\ref{rho1}) into Friedmann equation
(\ref{Friedeq02}) we can
obtain the differential equation for the Hubble parameter as\\
\begin{equation}\label{Friedeq03}
H^2(1-\alpha)-\frac{\epsilon}{r_c}H-\beta \dot{H}=0.
\end{equation}
Solving this equation, the Hubble parameter is obtained as
\begin{equation}\label{Hubble1}
H(t)=\frac{\epsilon}{r_{\rm c}(1-\alpha)+c_1\epsilon
e^{\frac{\epsilon t}{\beta r_{\rm c}}}},
\end{equation}
where $c_1$ is constant of integration. Since for $r_{\rm c}\gg1$,
the effects of the extra dimension should be disappeared and the
result of \cite{Granda2}, namely
\begin{equation}\label{Hubble}
H(t)=\frac{\beta}{\alpha-1}\frac{1}{t},
\end{equation}
must be restored, thus the constant $c_1$ should be chosen as
\begin{equation}
c_1=\frac{r_{\rm c}(\alpha-1)}{\epsilon}.
\end{equation}
Substituting $c_1$ in Eq. (\ref{Hubble1}), we obtain
\begin{equation}\label{Hubble2}
H(t)=\frac{\epsilon}{r_{\rm c}(1-\alpha)(1-e^{\frac{\epsilon
t}{\beta r_{c}}})}.
\end{equation}
Taking the time derivative of Eq. (\ref{Hubble2}) yields
\begin{equation}\label{Hubbledot}
\dot{H}(t)=\frac{1}{\beta r_{\rm
c}^2(1-\alpha)}\frac{e^{\frac{\epsilon t}{\beta
r_{c}}}}{(1-e^{\frac{\epsilon t}{\beta r_{c}}})^{2}}.
\end{equation}
We can also solve Eq. (\ref{Hubble2}) to obtain the scale factor.
We find
\begin{equation}\label{scale}
a(t)=a_0 \Big(e^{\frac{-\epsilon t}{\beta
r_c}}-1\Big)^{\frac{\beta}{\alpha-1}}.
\end{equation}
Using the fact that $1+z=a_{0}/a$, where $z$ is the redshift
parameter, and combining Eqs. (\ref{Hubble2}) and (\ref{scale}) we
find explicitly the Hubble parameter as a function of $z$,
\begin{eqnarray}\label{Hubble3}
H(z)&=&\frac{\epsilon}{r_{\rm
c}(1-\alpha)}\frac{\left[1+(1+z)^{\frac{1-\alpha}{\beta}}\right]}{(1+z)^{\frac{1-\alpha}{\beta}}}\nonumber \\
&=&\frac{\epsilon}{r_{\rm
c}(1-\alpha)}\left[1+(1+z)^{\frac{\alpha-1}{\beta}}\right].
\end{eqnarray}
This equation is valid in two cases. First, for $\epsilon=+1$ and
$\alpha<1$, and second for $\epsilon=-1$ and $\alpha>1$, and has
no solution for $\alpha=1$. Inserting Eqs. (\ref{Hubble2}) and
(\ref{Hubbledot}) in (\ref{rho1}), we get
\begin{equation}\label{rho}
\rho_{D}=\frac{3}{r_{c}^2(1-\alpha)^2}
\frac{\alpha+(1-\alpha)e^{\frac{\epsilon t}{\beta
r_{c}}}}{(1-e^{\frac{\epsilon t}{\beta r_{c}}})^{2}}.
\end{equation}
Substituting Eqs. (\ref{Hubble2}) and (\ref{Hubbledot}) in the
time derivative of Eq. (\ref{rho1}),
\begin{equation}\label{HDEdot}
\dot{\rho}_{D}=3(2\alpha\dot{H}H+\beta\ddot{H}),
\end{equation}
we arrive at
\begin{equation}\label{rhodot}
\dot{\rho}_D=\frac{3\epsilon e^{\frac{\epsilon t}{\beta
r_{c}}}}{\beta r_{c}^3(1-\alpha)^2(1-e^{\frac{\epsilon t}{\beta
r_{c}}})^2}\Big[\frac{2(\alpha+(1-\alpha)e^{\frac{\epsilon t}{\beta
r_{c}}})}{(1-e^{\frac{\epsilon t}{\beta r_{c}}})}+1-\alpha\Big].
\end{equation}
Combining  Eqs. (\ref{Hubble2}), (\ref{rho}) and (\ref{rhodot})
with (\ref{Conserveq}), we find the equation of state parameter of
HDE on the brane as a function of time,
\begin{equation}\label{wDt}
\omega_D(t)=-1-\frac{(1-\alpha)e^\frac{\epsilon t}{\beta
r_{c}}}{3\beta}\Big[1+\frac{1}{\alpha+(1-\alpha)e^\frac{\epsilon
t}{\beta r_{c}}}\Big].
\end{equation}
In order to see the evolution of $\omega_D$ during the history of the
universe, it is better to express it as a function of the redshift
parameter $z$,
\begin{equation}\label{wDz}
\omega_D(z)=-1-\frac{1-\alpha}{3\beta}
\Big[\frac{(1+\alpha)(1+z)^{\frac{1-\alpha}{\beta}}+2}
{(1+(1+z)^{\frac{1-\alpha}{\beta}})(1+\alpha(1+z)^{\frac{1-\alpha}{\beta}})}\Big].
\end{equation}
\begin{figure}[htp]
\begin{center}
\includegraphics[width=8cm]{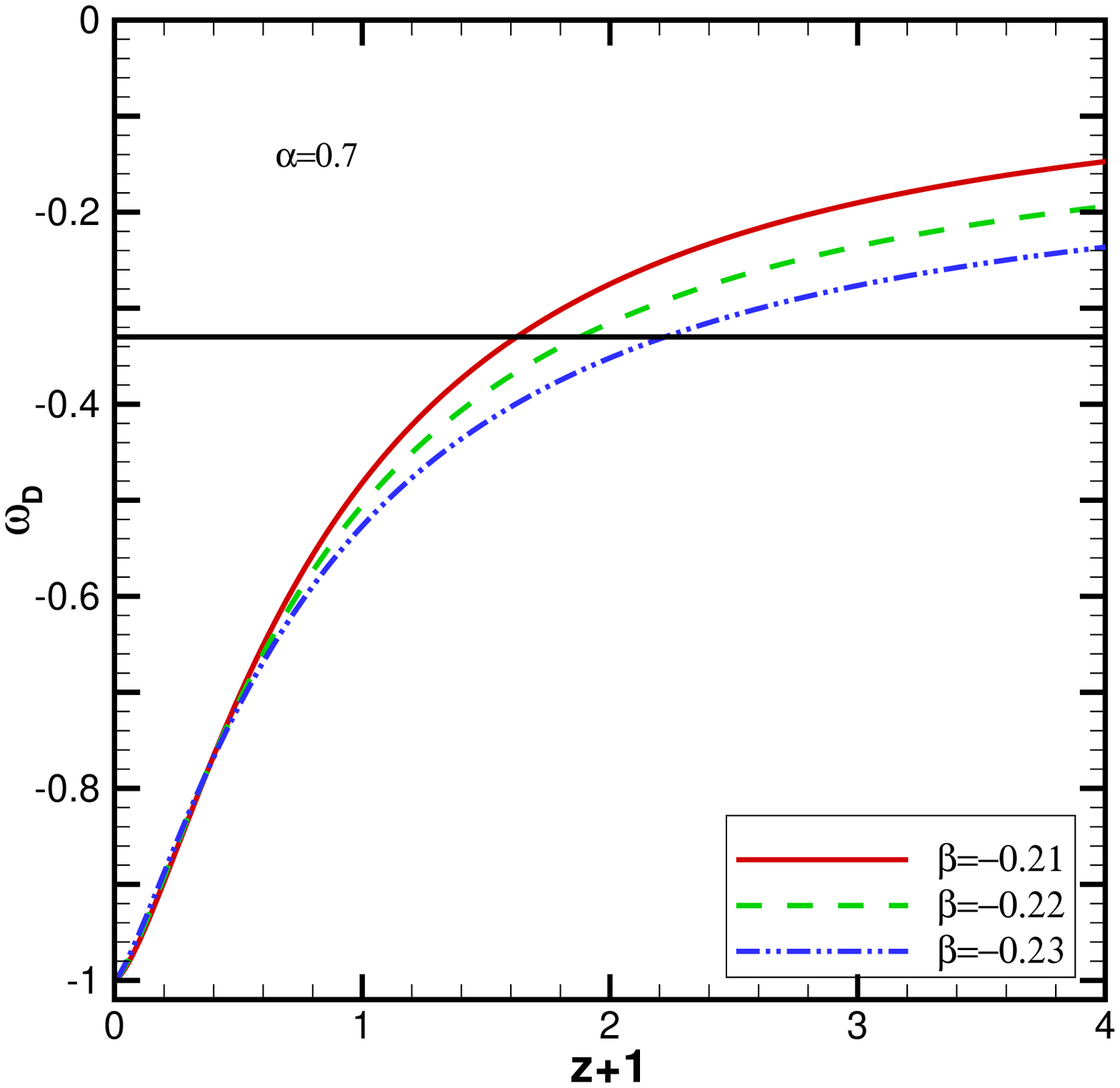}
\includegraphics[width=8cm]{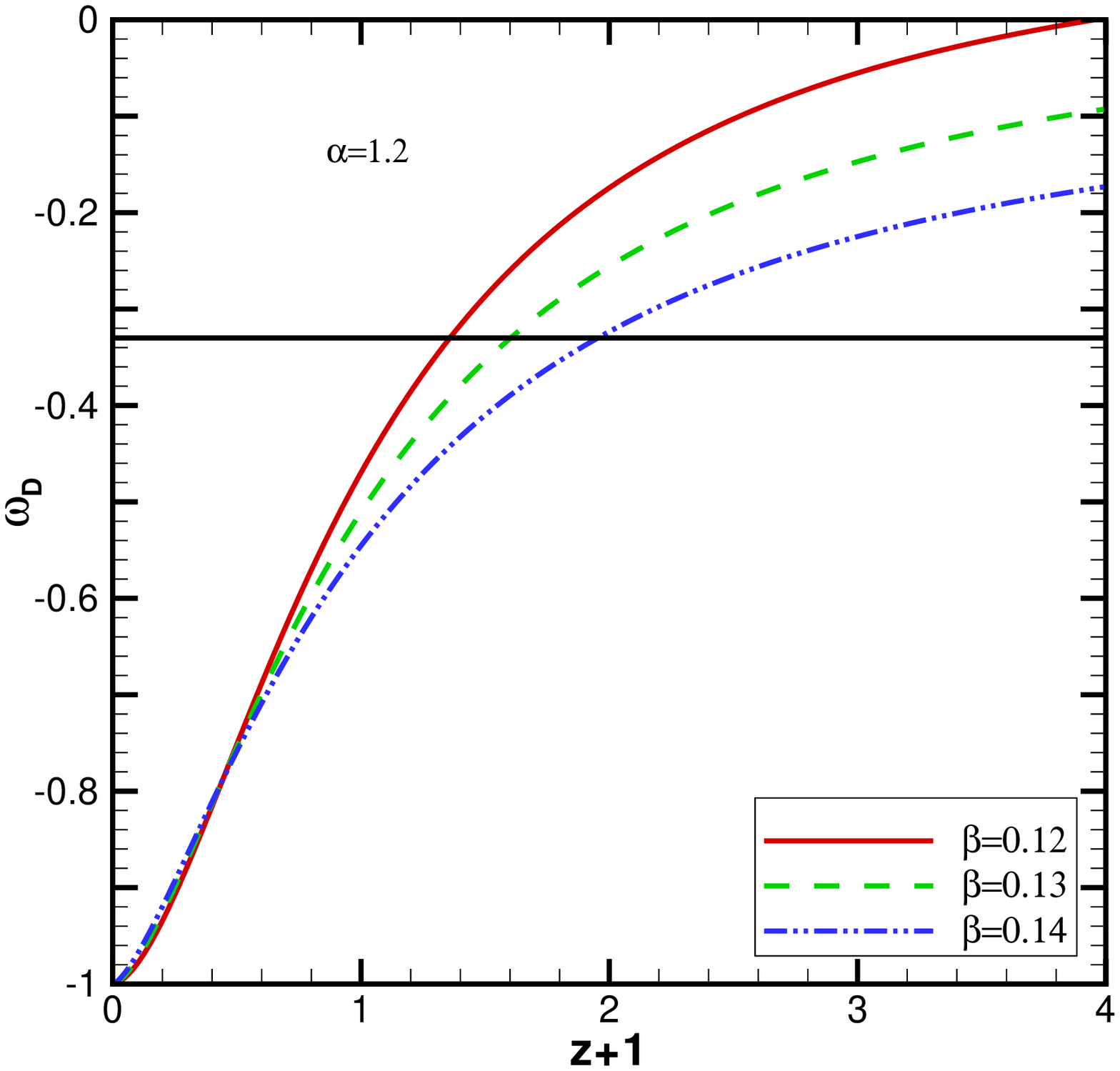}
\caption{The evolution of  $\omega_{D}$ versus redshift parameter $z$
for different model parameters. Left panel corresponds to
$\alpha<1$ and $\beta<0$, while the right panel shows the case
$\alpha>1$ and $\beta>0$. }\label{w}
\end{center}
\end{figure}

To check the limit of (\ref{wDz}) in standard cosmology where
$r_{c}\gg1$, we note that, from Eq. (\ref{scale}) and relation
$a/a_0=(1+z)^{-1}$ we have  $
(a/a_0)^{\frac{\alpha-1}{\beta}}=(1+z)^{\frac{1-\alpha}{\beta}}\rightarrow0$,
as $r_{c}\gg1$. Therefore, (\ref{wDz}) reduces to
\begin{equation}\label{wz2}
\omega_D=-1+\frac{2}{3}\frac{\alpha-1}{\beta},
\end{equation}
which is exactly the result obtained in \cite{Granda2}. It is
worth mentioning that unlike in standard cosmology, where the
equation of state parameter of HDE with GO cutoff becomes a
constant \cite{Granda2}, in DGP braneworld scenario $\omega_D$ varies
with time. Thus, it seems that the presence of the extra dimension
brings rich physics. For $\alpha=1$ we have $\omega_D=-1$, similar to
the cosmological constant. Let us consider the case with
$\alpha<1$ and $\alpha>1$ separately. In the first case where
$\alpha<1$, we find that the equation of state parameter can
explain the acceleration of the universe provided  $\beta<0$. In the
second case where $\alpha>1$, the accelerated expansion can be
achieved for $\beta>0$. In both cases, our universe has a
transition from deceleration to the acceleration phase around
$0.4\leq z\leq1$, compatible with the recent observations
\cite{Daly,Kom1,Kom2}, and mimics the cosmological constant at the
late time. These results can be easily seen in Fig. \ref{w}
which shows the behaviour of $\omega_D$ versus $z$ for both cases.

\begin{figure}[htp]
\begin{center}
\includegraphics[width=8cm]{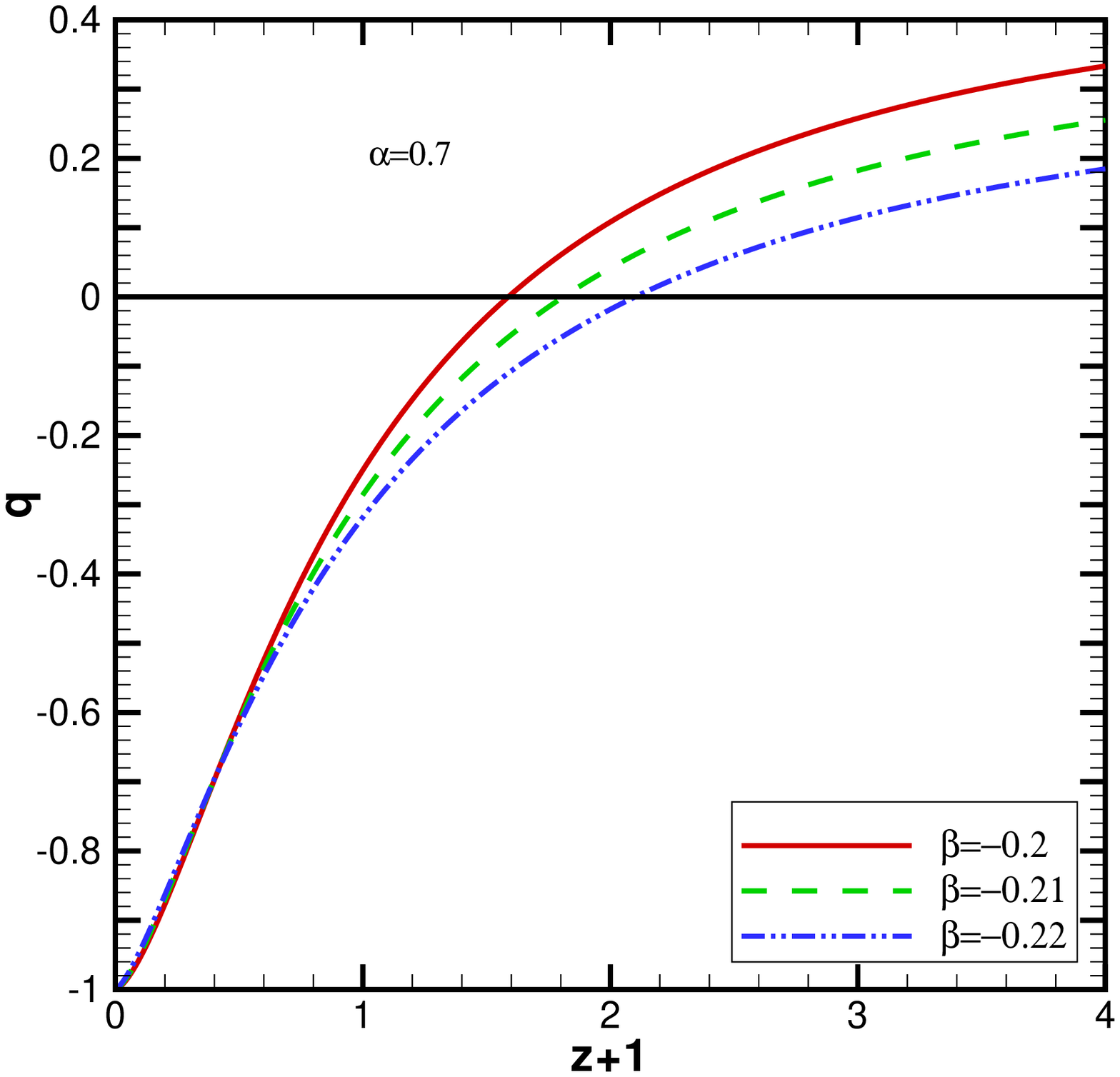}
\includegraphics[width=8cm]{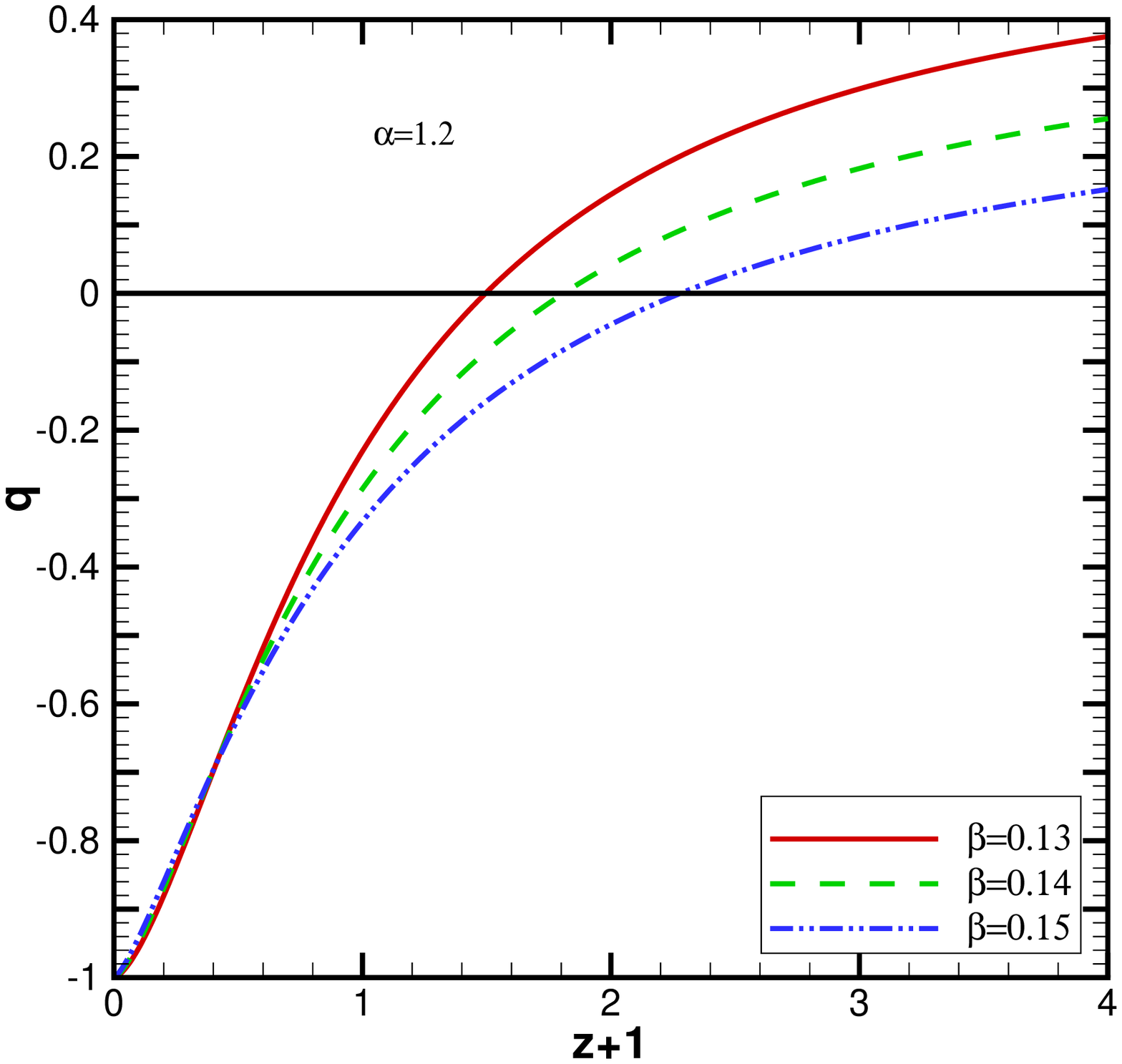}
\caption{The evolution of the deceleration parameter $q$ versus
redshift parameter $z$ for different model parameters. Left panel
corresponds to $\alpha<1$ and $\beta<0$, while the right panel
shows the case $\alpha>1$ and $\beta>0$. }\label{q}
\end{center}
\end{figure}

The first and second derivatives of the distance can be combined
to obtain the deceleration parameter $q$. It was shown that the
zero redshift value of $q_{0}$, is independent of space curvature,
and can be obtained from the first and second derivatives of the
coordinate distance \cite{Daly}. It was argued that $q_{0}$, which
indicates whether the universe is accelerating at the current
epoch, can be obtained directly from the supernova and radio
galaxy data \cite{Daly}. The deceleration parameter is given by
\begin{equation}
q=-1-\frac{\dot{H}}{H^2}.
\end{equation}
Using Eqs. (\ref{Hubble2}) and (\ref{Hubbledot}), the deceleration
parameter is obtained as
\begin{equation}\label{q1}
q=-1-\frac{1-\alpha}{\beta\left[1+(1+z)^\frac{1-\alpha}{\beta}\right]}.
\end{equation}
For $\alpha=1$ we have $q=-1$. Although, the equation of state and
the deceleration parameters do not depend  explicitly on the
crossover length scale $r_{\rm c}$ which is the characterization
of the DGP branworld, they depend on $r_{\rm c}$ via the relation
between the scale factor and the redshift parameter as in (\ref{scale}).
We have plotted the behavior of the deceleration parameter versus
$z$ in figure \ref{q}. Form this figures, we see that the
transition from deceleration phase to the acceleration phase can
be occurred around $0.4\leq z\leq1$, which is consistent with
the recent observations \cite{Daly}. The zero-redshift value of
the deceleration parameter is obtained as
\begin{equation}\label{q2}
q=-1+\frac{\alpha-1}{2\beta}.
\end{equation}
Recent cosmological data from the combined sample of $192$
supernovae and $30$ radio galaxies, show that $q_0=-0.30\pm 0.18$,
for a window function of width $0.4$ in redshift  and a transition
from deceleration to the acceleration phase at redshift $z=
0.78\pm 0.37$ \cite{Daly}. The zero-redshift value (\ref{q2}), for
$\alpha=0.7$ and $\beta=-0.22$, leads to $q_0=-0.32$, while for
$\alpha=1.2$ and $\beta=0.15$, we get $q_0=-0.33$, which is
consistent with the observations \cite{Daly}.

\section{Correspondence with scalar field models \label{SF}}
The dynamical DE proposal is often realized by some scalar field
mechanism which suggests that the energy form with negative
pressure is provided by a scalar field evolving down a proper
potential. Scalar fields naturally arise in particle physics
including supersymmetric field theories and string/M theory.
Therefore, scalar field is expected to reveal the dynamical
mechanism and the nature of DE. However, although fundamental
theories such as string/M theory do provide a number of possible
candidates for scalar fields, they do not predict its potential,
$V(\phi)$, uniquely. Consequently, it is meaningful to reconstruct
the potential $V(\phi)$  from some DE models possessing some
significant features of the quantum gravity theory, such as HDE
models. Famous examples of scalar field DE models include
quintessence \cite{Peebles}, K-essence \cite{Armendariz}, tachyon
\cite{Sen}, phantom \cite{Caldwell}, ghost condensate
\cite{Arkani,Piazza}, quintom \cite{Feng}, and so forth. For a
comprehensive review on scalar filed models of dark energy, see
\cite{cop,Li2011}. Generically, there are two points of view on
the scalar field models of dynamical DE. One viewpoint regards the
scalar field as a fundamental field of the nature. The nature of
DE is, according to this viewpoint, completely attributed to some
fundamental scalar field which is omnipresent in supersymmetric
field theories and in string/M theory. The other viewpoint
supports that the scalar field model is an effective description
of an underlying theory of DE. If we regard the scalar field model
as an effective description of such a theory, we should be capable
of using the scalar field model to mimic the evolving behavior of
the HDE and reconstructing the scalar field model according to the
evolutionary behavior of HDE.

In this section we implement a correspondence between HDE in DGP
braneworld with GO cutoff and various scalar field models, by
equating the equation of state parameters for these models with the obtained
equation of state parameter of Eq. (\ref{wDz}).
\subsection{Reconstructing holographic quintessence model}
Let us start with reconstructing the potential and dynamics of
quintessence scaler field. The energy density and pressure of the
quintessence scalar field are given by
\begin{eqnarray}\label{rhophi}
\rho_\phi=\frac{1}{2}\dot{\phi}^2+V(\phi),\\
p_\phi=\frac{1}{2}\dot{\phi}^2-V(\phi). \label{pphi}
\end{eqnarray}
Thus the potential and the kinetic energy term can be written as
\begin{eqnarray}\label{vphi}
&&V(\phi)=\frac{1-\omega_\phi}{2}\rho_{\phi},\\
&&\dot{\phi}^2=(1+\omega_\phi)\rho_\phi. \label{dotphi}
\end{eqnarray}
where $\omega_{\phi}=p_{\phi}/\rho_{\phi}$. In order to implement the
correspondence between HDE and quintessence scaler field, we
identify $\rho_\phi=\rho_D$ and $\omega_\phi=\omega_D$. Using Eqs.
(\ref{wDz}), (\ref{vphi}) and (\ref{dotphi}), we find
\begin{equation}\label{phidotq}
\dot{\phi}^2=\frac{1}{\beta
r_{c}^2(\alpha-1)}\frac{(1+\alpha)(1+z)^{\frac{1-\alpha}{\beta}}+2}{(1+z)^{\frac{2(1-\alpha)}{\beta}}},
\end{equation}
\begin{equation}\label{Vphiq}
V(z)=\frac{3}{2r_{c}^2}\frac{(1+z)^{\frac{2(\alpha-1)}{\beta}}}{(1-\alpha)^2}\Big[\left(2\alpha(1+z)^{\frac{1-\alpha}{\beta}}+2\right)
\left[(1+z)^{\frac{1-\alpha}{\beta}}+1\right]+
\frac{1-\alpha}{3\beta}\left((1+\alpha)(1+z)^{\frac{1-\alpha}{\beta}}+2\right)\Big].
\end{equation}
Eq. (\ref{phidotq}) can also be rewritten
\begin{equation}
\frac{d\phi}{d\ln a}=H^{-1}\dot{\phi}=\frac{1-\alpha}{
\epsilon\sqrt{\beta(\alpha-1)}}\frac{\sqrt{(1+\alpha)(1+z)^{\frac{1-\alpha}{\beta}}+2}}{1+(1+z)^{\frac{1-\alpha}{\beta}}}.
\end{equation}
Using relation $a=a_0(1+z)^{-1}$ one can  also obtain
\begin{equation}
\frac{d\phi}{dz}=\frac{1}{\epsilon(1+z)}\sqrt{\frac{\alpha-1}{\beta}}
\frac{\sqrt{(1+\alpha)(1+z)^{\frac{1-\alpha}{\beta}}+2}}{1+(1+z)^{\frac{1-\alpha}{\beta}}}.
\end{equation}
Integrating, yields
\begin{equation}\label{phizq}
\phi(z)=\int{\frac{dz}{\epsilon(1+z)}\sqrt{\frac{\alpha-1}{\beta}}
\frac{\sqrt{(1+\alpha)(1+z)^{\frac{1-\alpha}{\beta}}+2}}{1+(1+z)^{\frac{1-\alpha}{\beta}}}}.
\end{equation}
\begin{figure}[htp]
\begin{center}
\includegraphics[width=8cm]{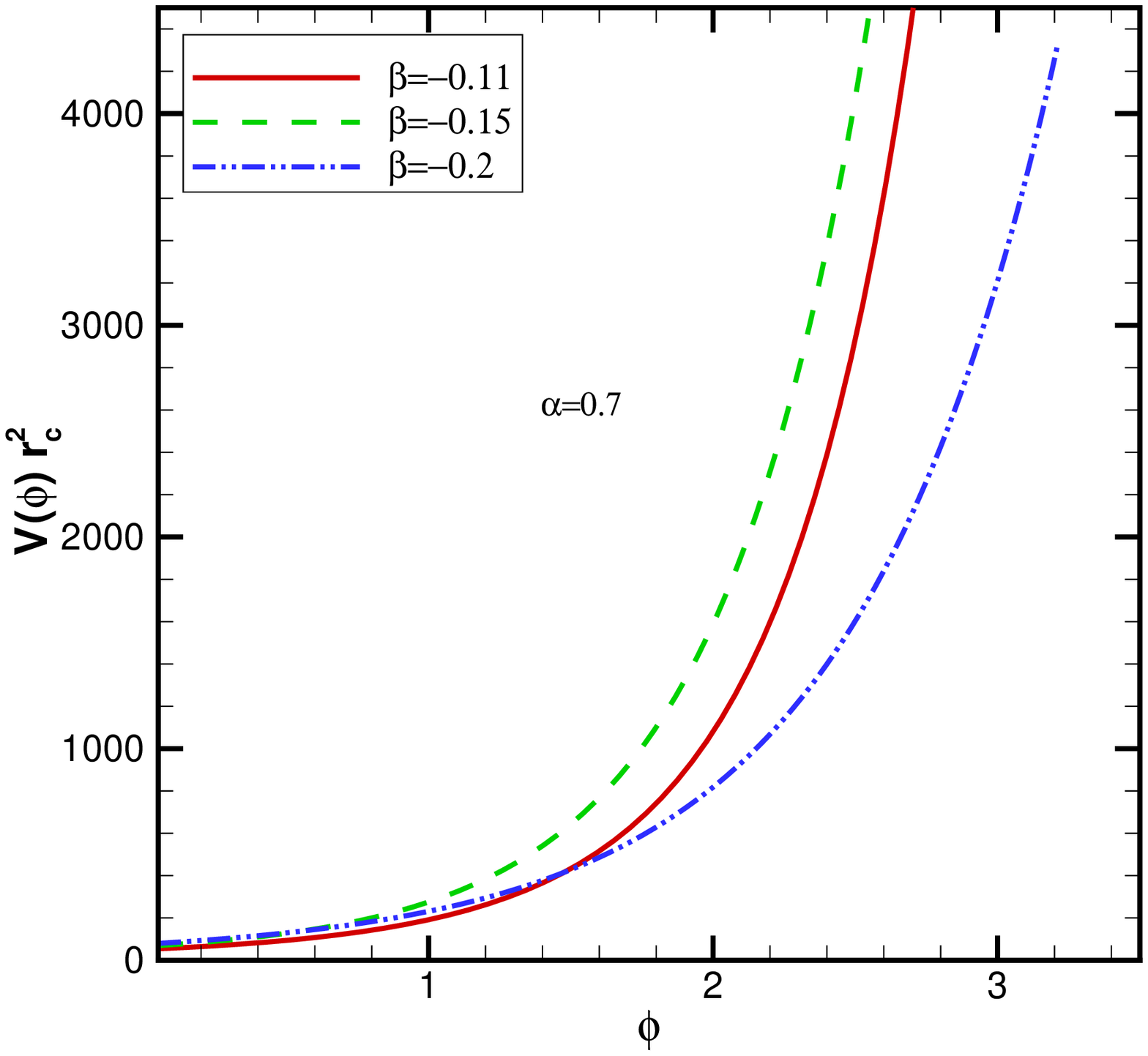}
\includegraphics[width=8cm]{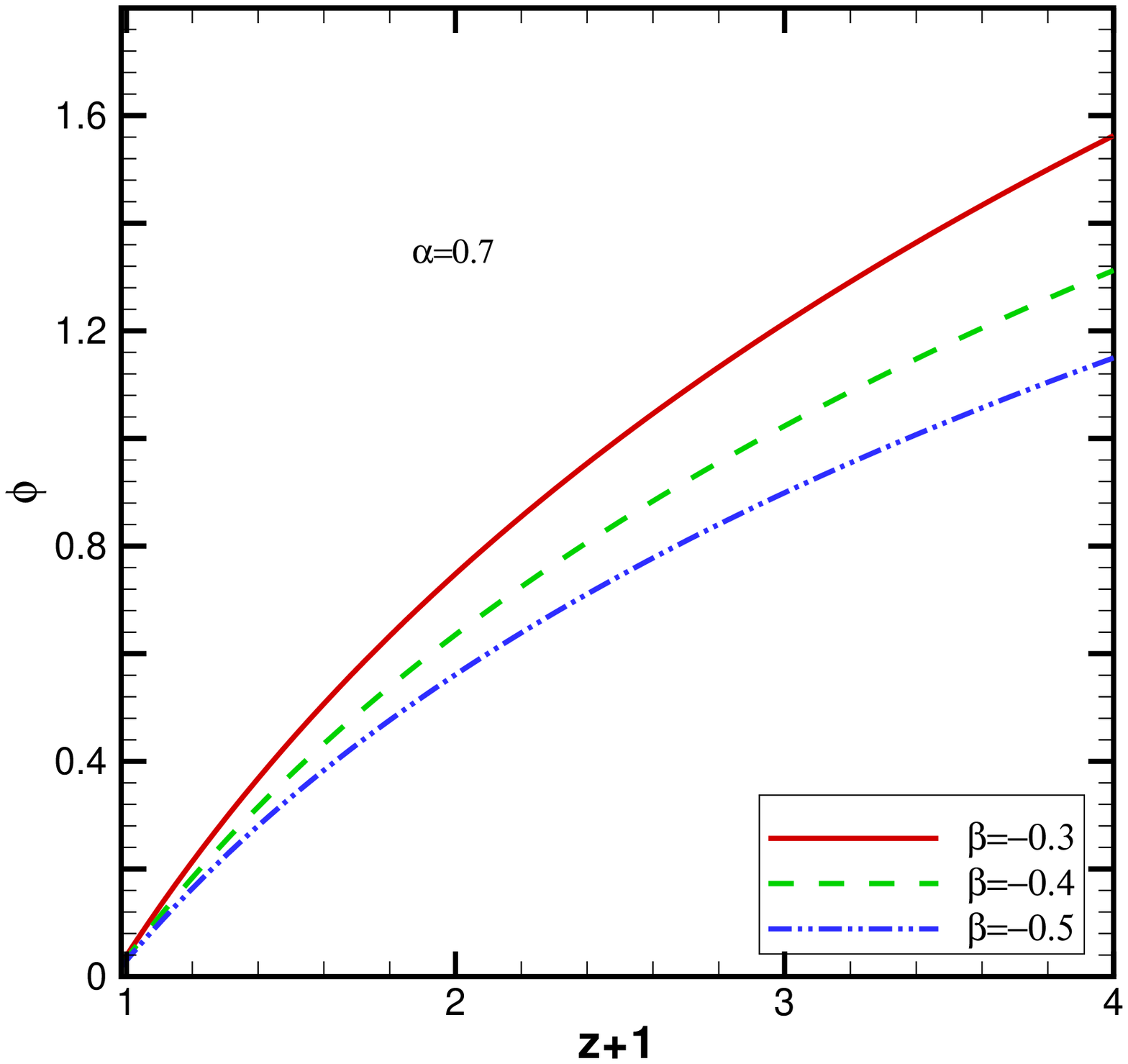}
\caption{Left panel shows the reconstruction of the quintessence
potential $V(\phi)$ for different parameters. In the right panel
the behavior  of the  quintessence scalar field $\phi(z)$ for HDE
model in DGP braneworld is illustrated for different
parameters.}\label{quint}
\end{center}
\end{figure}
Therefore, we have established holographic quintessence DE model
and reconstructed the potential and the dynamics of scalar field
according to evolutionary  behaviour of HDE on the brane.
Theoretically, one may expect to omit the redshift $z$ from
(\ref{Vphiq}) and (\ref{phizq}) to derive the potential as a
function of the scalar field, namely, $V(\phi)$. However, due to
the complexity of the equations, the analytical form of the
potential $V=V(\phi)$ is hard to be derived. Nevertheless, we can
plot the reconstructed potential as a function of $\phi$
numerically. The reconstructed quintessence potential $V(\phi)$
and the evolutionary form of the field are plotted in figure
\ref{quint}, where we have taken the zero redshift value of the
scalar field equal to zero, namely, $\phi(z=0)=0$. Selected curves
are plotted for $\alpha<1$ and the different model parameter
$\beta<0$. We can also plot the figures for the case $\alpha>1$
and $\beta>0$, however, the behaviour is the same as the former
case and for the economic reason we have not plotted here the
latter case. From these figures we can see the dynamics of the
potential as well as the scalar field explicitly. Figure
\ref{quint} shows that the reconstructed quintessence potential is
steeper in the early epoch and becomes flat near the present time.
In other words, it mimics a cosmological constant at the the
present time.
\subsection{Reconstructing holographic tachyon model}
The tachyon field is another candidate for DE. The equation of
state parameter of the rolling tachyon smoothly interpolates
between $-1$ and $0$ \cite{Gibbons}. Thus, tachyon can be realized
as a suitable candidate for the inflation at high energy
\cite{Mazumdar} as well as a source of DE depending on the form of
the tachyon potential \cite{Padmanabhan}. Choosing different
self-interacting potentials in the tachyon field model lead to
different consequences for the resulting DE model. Due to all the above
reasons, the reconstruction of tachyon potential $V(\phi)$ is of
great importance. The correspondence between tachyon field and
various DE scenarios such as HDE \cite{Setare} and agegraphic DE
\cite{tachADE} has been already established. The study has also
been generalized to the entropy corrected holographic and
agegraphic DE models \cite{ecde}.

The tachyon condensates in a class of string theories and can be
described by an effective scalar field with a Lagrangian of the
form by\cite{Bergshoeff}
\begin{equation}
L=-V(\phi)\sqrt{1-g^{\mu\nu}\partial_{\mu}\phi\partial_{\nu}\phi},
\end{equation}
where $V(\phi)$ is the tachyon potential. The corresponding energy
momentum tensor for the tachyon field can be written in a perfect
fluid form
\begin{equation}
T_{\mu\nu}=(p+\rho)u_{\mu}u_{\nu}-pg_{\mu\nu}
\end{equation}
where $\rho$ and $p$ are, respectively, the energy density and
pressure of the tachyon, and the velocity $u_{\mu}$ is
\begin{equation}
u_{\mu}=\frac{\partial_{\mu}\phi}{\sqrt{\partial_{\nu}\phi\partial^{\nu}\phi}}
\end{equation}
The energy density and pressure of tachyon field are given by
\begin{eqnarray}\label{rhot}
\rho&=&-T^{0}_{0}=\frac{V(\phi)}{\sqrt{1-\dot{\phi}^2}},\\
p&=&T^{i}_{i}=V(\phi)\sqrt{(1-\dot{\phi}^2)}.\label{pt}
\end{eqnarray}
Thus the equation of state parameter of tachyon field is given by
\begin{equation}\label{wT}
\omega_T=\frac{p}{\rho}=\dot\phi^2-1.
\end{equation}
To establish the correspondence between HDE and tachyon field, we
equate  $\omega_D$ with $\omega_T$. Combining Eqs. (\ref{wDz}), (\ref{rhot})
and (\ref{pt}), we  find
\begin{eqnarray}\label{Vt1}
V(z)&=&\frac{3}{r_c^2(1-\alpha)^2}\sqrt{1-\frac{\alpha-1}{3\beta}
\left[\frac{(1+\alpha)(1+z)^{\frac{1-\alpha}{\beta}}+2}{(1+(1+z)^{\frac{1-\alpha}{\beta}})(1+\alpha
(1+z)^{\frac{1-\alpha}{\beta}})}\right]}\frac{(1+\alpha(1+z)^{\frac{1-\alpha}{\beta}})(1+(1+z)^{\frac{1-\alpha}{\beta}})}{(1+z)^{\frac{2(1-\alpha)}{\beta}}},\\
\dot{\phi}^2(z)&=&\frac{\alpha-1}{3\beta}
\left[\frac{(1+\alpha)(1+z)^{\frac{1-\alpha}{\beta}}+2}{(1+(1+z)^{\frac{1-\alpha}{\beta}})(1+\alpha
(1+z)^{\frac{1-\alpha}{\beta}})}\right].\label{phit1}
\end{eqnarray}
We can further rewrite (\ref{phit1}) as
\begin{equation}\label{phit2}
\frac{d\phi}{d (\ln
a)}=H^{-1}\dot{\phi}=\frac{-r_c(1+z)^{\frac{1-\alpha}{\beta}}}{\epsilon}\sqrt{\frac{(\alpha-1)^3}{3\beta}}
\left[\frac{(1+\alpha)(1+z)^{\frac{1-\alpha}{\beta}}+2}{(1+(1+z)^{\frac{1-\alpha}{\beta}})^3(1+\alpha
(1+z)^{\frac{1-\alpha}{\beta}})}\right]^{\frac{1}{2}}.
\end{equation}
Using the fact that
\begin{equation}
\frac{d\phi}{dz}=\frac{d\phi}{da}\frac{da}{dz}=-(1+z)^{-1}\frac{d\phi}{d(\ln a)},
\end{equation}
we have
\begin{equation}\label{phit3}
\frac{d\phi}{dz}=\frac{r_c(1+z)^{\frac{1-\alpha}{\beta}}}{\epsilon(1+z)}\sqrt{\frac{(\alpha-1)^3}{3\beta}}
\left[\frac{(1+\alpha)(1+z)^{\frac{1-\alpha}{\beta}}+2}{(1+(1+z)^{\frac{1-\alpha}{\beta}})^3(1+\alpha
(1+z)^{\frac{1-\alpha}{\beta}})}\right]^{\frac{1}{2}}.
\end{equation}
Integrating and setting the constant of integration equal to zero
by assuming $\phi(z=0)=0$, we obtain
\begin{equation}\label{phit4}
\phi(z)=\int{\frac{r_c(1+z)^{\frac{1-\alpha}{\beta}}}{\epsilon(1+z)}\sqrt{\frac{(\alpha-1)^3}{3\beta}}
\left[\frac{(1+\alpha)(1+z)^{\frac{1-\alpha}{\beta}}+2}{(1+(1+z)^{\frac{1-\alpha}{\beta}})^3(1+\alpha
(1+z)^{\frac{1-\alpha}{\beta}})}\right]^{\frac{1}{2}}dz}.
\end{equation}
\begin{figure}[htp]
\begin{center}
\includegraphics[width=8cm]{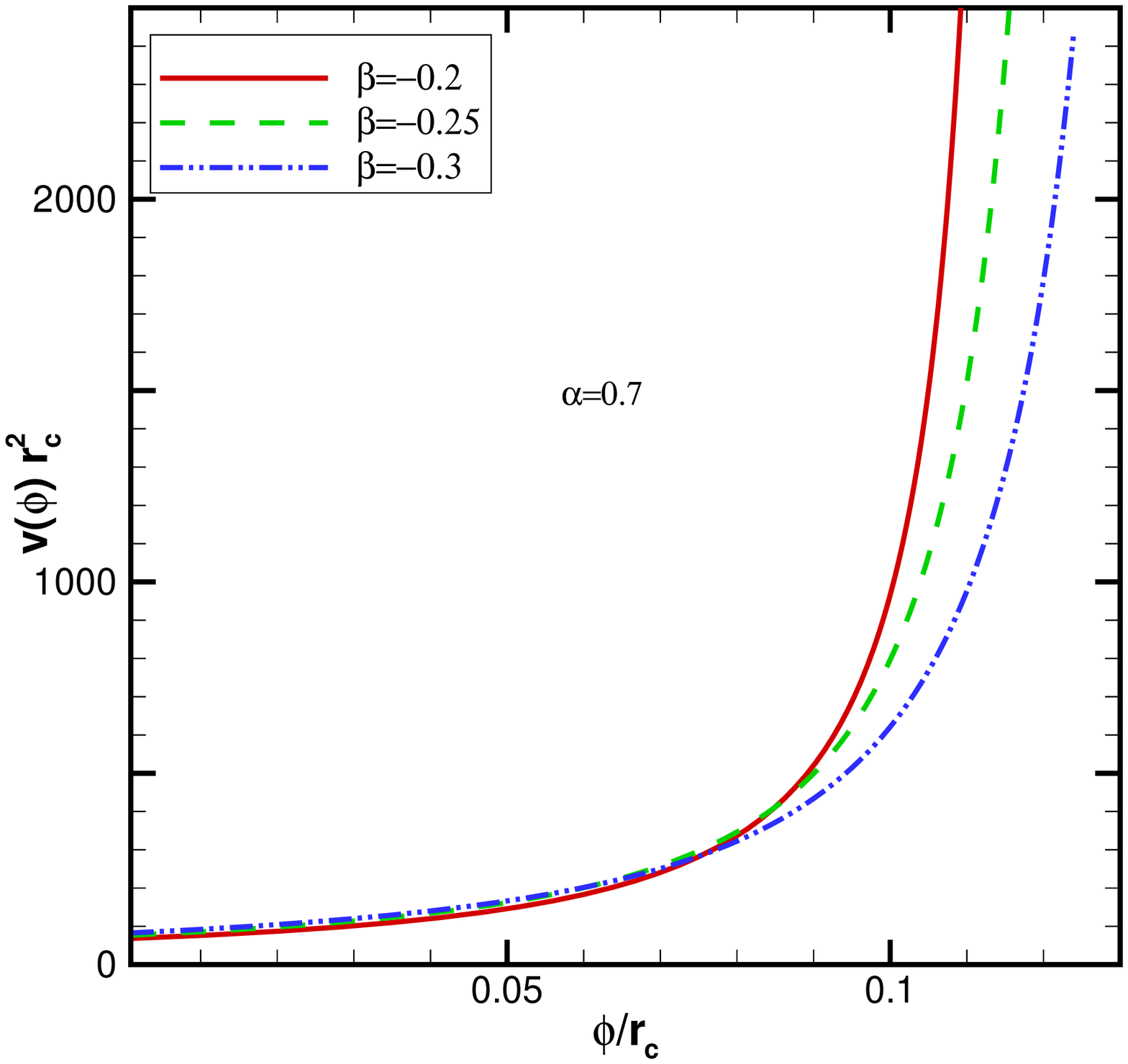}
\includegraphics[width=8cm]{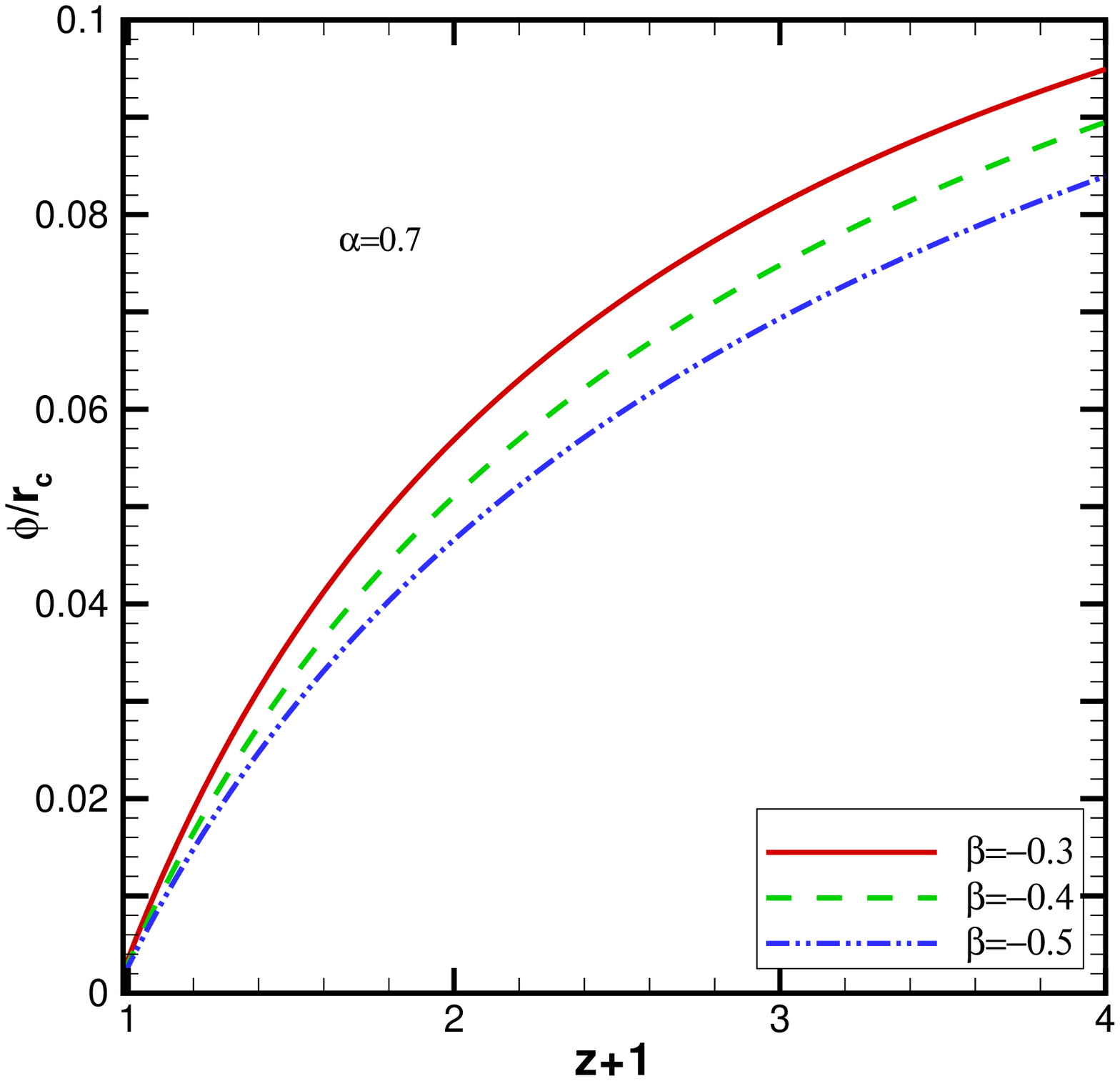}
\caption{Left panel shows the reconstruction of the tachyon
potential $V(\phi)$ for different parameters. In the right panel
the behavior  of the  tachyon field $\phi(z)$ for HDE model in DGP
braneworld is illustrated for different
parameters.}\label{Tachyon}
\end{center}
\end{figure}
In this way we connect the HDE in DGP braneworld with a tachyon
field, and reconstruct the potential and the dynamics of the
tachyon field which describe tachyon cosmology. The reconstructed
tachyon field and the evolution of the tachyon potential are
plotted in figure \ref{Tachyon}. Again, we see that tachyon
potential is steeper in the early epoch and becomes flat near
today. Thus, the universe enters a de-Sitter phase at the late
time and the constant potential plays the role of cosmological
constant.

\section{stability of the model }
In this section we would like to investigate the stability of the
HDE in DGP braneworld  against perturbation. It is expected that
any viable DE model should result a stable DE dominated universe.
A simple, but not complete, approach to check the stability of a
proposed DE model is to study the behavior of the square sound
speed ($v_s^2 = dp/d\rho$) \cite{peeb}. It was argued that the
sign of $v_s^2$ plays a crucial role in determining the stability
of the background evolution. If $v_s^2<0$, it implies that the
perturbation of the background energy density is not an
oscillatory function and may grow or decay with time, and so we
have the classical instability of a given perturbation. On the
other hand, the positivity of $v_s^2>0$  indicates that the perturbation in the
energy density, propagates in the environment and so we expect a
stable universe against perturbations. It is important to note
that the positivity of $v_s^2$ is necessary but is not enough to conclude that the
model is stable. Indeed, the negativity of it shows a sign of instability in the
model. The behavior of the squared sound speed for HDE
\cite{Myung}, agegraphic DE \cite{Kim} and the ghost DE model
\cite{ebrins,saaidi} were investigated. It was found that all
these models \cite{Myung,Kim,ebrins,saaidi} are instable against
background perturbations and so cannot lead to a stable DE
dominated universe.

In the linear perturbation regime, the perturbed energy density of
the background can be written as
\begin{equation}\label{pert1}
\rho(t,x)=\rho(t)+\delta\rho(t,x),
\end{equation}
where $\rho(t)$ is unperturbed background energy density. The
energy conservation equation ($\nabla_{\mu}T^{\mu\nu}=0$) yields
\cite{peeb}
\begin{equation}\label{pert2}
\delta\ddot{\rho}=v_s^2\nabla^2\delta\rho(t,x).
\end{equation}
For $v^2_s>0$, Eq. (\ref{pert2}) becomes a regular wave equation
whose solution is given by $\delta\rho=
\delta\rho_{0}e^{-i\omega_{0} t+ik.x}$, which indicates a
propagation mode for the density perturbations. For $v^2_s<0$, the
perturbation becomes an irregular wave equation and in this case
the frequency of the oscillations are pure imaginary and the
density perturbation will grow with time as $\delta\rho=
\delta\rho_{0}e^{\omega t+ik.x}$. Hence the negative squared speed
shows an exponentially growing mode for a density perturbation.
Thus the growing perturbation with time indicates a possible
emergency of instabilities in the background.
\begin{figure}[htp]
\begin{center}
\includegraphics[width=8cm]{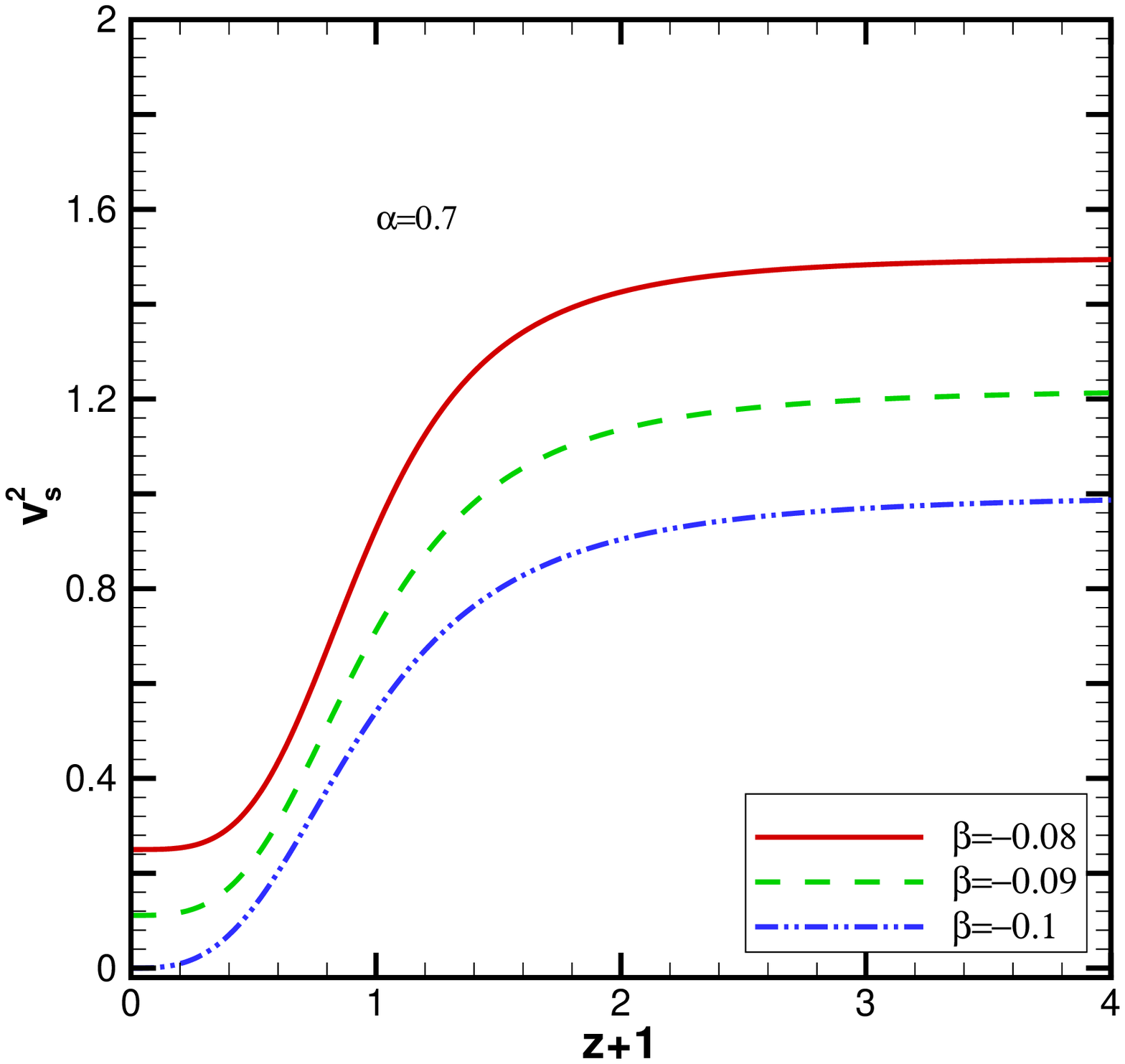}
\includegraphics[width=8cm]{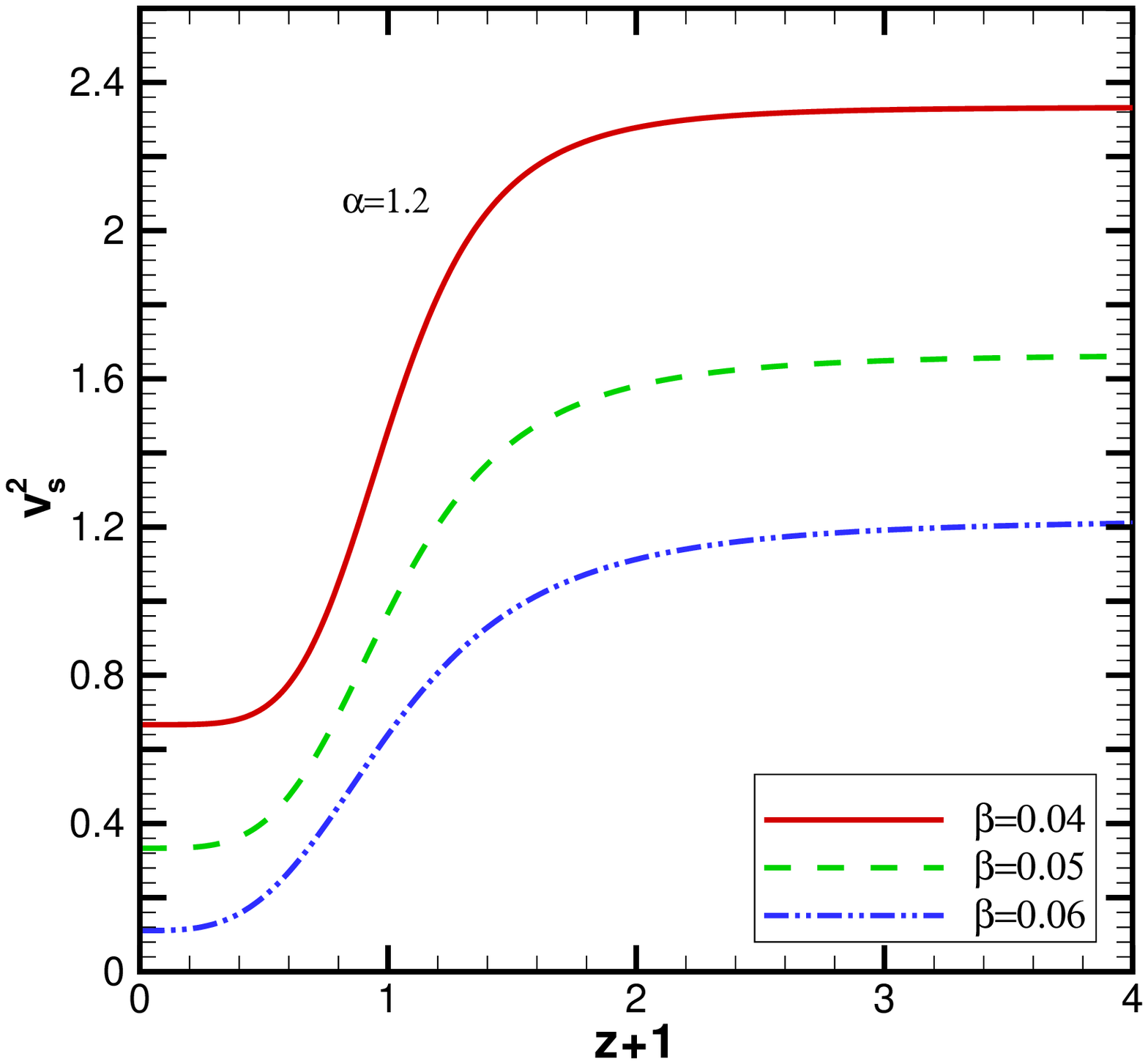}
\caption{The behaviour of $v_{s}^2$ for HDE model in DGP
braneworld for different model parameters. Left panel corresponds
to $\alpha<1$ and $\beta<0$, while the right panel shows the case
$\alpha>1$ and $\beta>0$. }\label{Vs}
\end{center}
\end{figure}
The squared speed of sound can be written as
\begin{equation}\label{vs1}
v^2_s=\frac{dp}{d\rho}=\frac{\dot{p}}{\dot{\rho}},
\end{equation}
where
\begin{equation}
\dot{p}=\dot{\omega}_D\rho+\omega_D\dot{\rho}.
\end{equation}
Taking the time derivative of (\ref{wDz}) and substituting the
result in (\ref{vs1}), after some calculation, we arrive at
\begin{equation}
v^2_s=\frac{\alpha-1}{3\beta}\Big[\frac{4+(1+\alpha)(1+z)^{\frac{1-\alpha}{\beta}}}
{2+(1+\alpha)(1+z)^{\frac{1-\alpha}{\beta}}}\Big]-1.
\end{equation}
The evolution of $v^2_s$ versus redshift parameter are plotted in
figure \ref{Vs}. From these figures we see that $v^2_s$ can be
positive provided the parameters of the model are chosen suitably.
For example, for $\alpha<1$, $\beta<0$, and also for $\alpha>1$,
$\beta>0$ the square of sound speed is always positive during the
history of the universe, and so in these cases the stable DE
dominated universe can be achieved.
\section{Concluding remarks}
In this paper, we have investigated the holographic model of DE in
the framework of DGP braneworld. We have chosen the GO IR cutoff
in the form, $L= (\alpha \dot{H}+\beta H^2)^{-1/2}$, with two free
parameters $\alpha$ and $\beta$, which should be constrained by
comparing with observations. This cutoff is the generalization of
the well-known Ricci scalar cutoff in flat universe. {We have
restricted  our study to the current cosmological epoch, and so we
have not considered the contributions from matter and radiation by
assuming that the dark energy $\rho_D$ dominates, thus the
Friedman equation becomes simpler. The main difference between the
HDE with GO cutoff studied in this work in the framework of DGP
braneword, with the one considered in standard cosmology
\cite{Granda2}, is that the equation of state parameter of the HDE
with GO cutoff in standard cosmology is a constant \cite{Granda2},
however, in DGP braneworld, due to the bulk effects, $w_D$ becomes
a time variable parameter.}  The application of the HDE with GO
cutoff, in DGP braneworld allows us to solve the Friedmann
equation and derive the Hubble parameter, $H(z)$, as well as the
scale factor, $a(t)$, analytically. We also obtained the equation
of state and the deceleration parameters of HDE as the functions
of the redshift parameter $z$ and plotted the evolutionary
behaviour of these parameters against $z$. The cosmological
quantities depend on the two model parameter $\alpha$ and $\beta$.
To show the viability of the model, we studied the zero-redshift
values of the cosmological parameters by chosen the suitable
values for $\alpha$ and $\beta$. For example, we found that for
$\alpha=0.7$ and $\beta=-0.21,-0.22,-0.23$, the transition from
deceleration phase to the acceleration phase occurred around
$0.4\leq z\leq 1$, which is consistent with recent observations
\cite{Daly}. The zero-redshift value of the deceleration parameter
was obtained $q_0=-0.32$ for $\alpha=0.7$, $\beta=-0.22$, and
$q_0=-0.33$ for $\alpha=1.2$, $\beta=0.15$ which is again
compatible with the cosmological data \cite{Daly}. Besides, for
$\alpha=1$ this model mimics a cosmological constant with
$\omega_D=-1$, and $q=-1$, independent of the redshift parameter
$z$. For $\alpha>1$, $\beta>0$ and $\alpha<1$, $\beta<0$ the
equation of state parameter, however, is always larger than $-1$,
and the Universe enters a de-Sitter phase at the late time.

We have also established a connection between the quintessence/tachyon
scalar field and the HDE in DGP braneworld. As a result,
we reconstructed the corresponding potentials of the scalar field,
$V(\phi)$, and the dynamics of the scalar fields as a function of
redshift parameter, $\phi=\phi(z)$, according to the evolutionary
behavior of the HDE model. Finally, we studied the stability of
the presented model by studying the evolution of the squared sound
speed  $v^2_s$ whose sign determines the sound stability of the
model. Interestingly enough, we found that $v^2_s>0$ provided the
parameters of the model are chosen suitably. For example, for
$\alpha=0.7$, $\beta=-0.1$, and $\alpha=1.2$, $\beta= 0.06$ the
squared sound speed is always positive during the history of the
universe, and so in this case the stable DE dominated universe can
be achieved. This is in contrast to HDE in standard cosmology,
which is instable against background perturbations and so cannot
lead to a stable DE dominated universe \cite{Myung}. This implies
that the presence of the extra dimension in HDE model, can bring
rich physics and in particular it has an important effect on the
stability of the HDE model. This issue deserves further
investigations.

\acknowledgments{We thank from the Research Council of Shiraz
University. This work has been supported financially by Research
Institute for Astronomy \& Astrophysics of Maragha (RIAAM), Iran.
}

\end{document}